\documentclass[twocolumn,aps,prb,showkeys,superscriptaddress,preprintnumbers]{revtex4-1}
\usepackage[latin9]{inputenc}
\usepackage{amsmath}
\usepackage{amssymb}
\usepackage{txfonts}
\usepackage[titletoc]{appendix}
\usepackage{array}
\usepackage{mathrsfs}
\usepackage{dcolumn}
\usepackage{graphicx}

\begin{document}

\title{Transport theory for electrical detection of the spin texture and
spin-momentum locking of topological surface states}
\author{Shi-Han Zheng, Hou-Jian Duan, Mou Yang, and Rui-Qiang Wang}
\email{rqwanggz@163.com}
\affiliation{Guangdong Provincial Key Laboratory of Quantum Engineering and Quantum
Material, ICMP and SPTE, South China Normal University, Guangzhou 510006,
China }
\date{\today }
\begin{abstract}
The surface states of three-dimensional topological insulators exhibit a
helical spin texture with spin locked to momentum. To date, however, the
direct all-electrical detection of the helical spin texture has remained
elusive owing to the lack of necessary spin-sensitive measurements. We here
provide a general theory for spin polarized transports of helical Dirac
electrons through spin-polarized scanning tunneling microscopy (STM). It is
found that different from conventional magnetic materials, the tunneling
conductance through the TI surface acquires an extra component determined by
the in-plane spin texture, exclusively associated with spin momentum
locking. Importantly, this extra conductance unconventionally depends on the
spatial azimuthal angle of the magnetized STM tip, which is never carried
out in previous STM theory. By magnetically doping to break the symmetry of
rotation and time reversal of the TI surface, we find that the measurement
of the spatial resolved conductance can reconstruct the helical structure of
spin texture. Furthermore, one can extract the SML angle if the in-plane
magnetization is induced purely by the spin-orbit coupling of surface Dirac
elections. Our theory offers an alternative way, rather than using angle
resolved photoemission spectroscopy, to electrical identify the helical spin
texture on TI surfaces.
\end{abstract}
\maketitle
The discovery of three-dimension topological insulators (TIs)\cite{Chen,Zhan}%
, such as Bi$_2$Se$_3$ and Bi$_2$Te$_3$, has currently triggered great
interest in surface electronics. The most striking hallmark of TIs is the
gapless topological surface states, which exhibits a spin texture with the
intrinsic spin of Dirac electrons locked to its momentum, thus protecting
the Dirac electrons immune to the backscattering off perturbations with time
reversal symmetry\cite{qi,Rous,bei}. This spin-momentum locking (SML) nature
provides a concept for electrical manipulation of spin in a controlled
fashion and makes TI particularly promising for spintronic devices and
topological quantum computation\cite{Hasa,moor,Fu}. Though such spin
helicity has been experimentally observed by spin-resolved angle resolved
photoemission spectroscopy (ARPES)\cite{hsi2,hsi,xia,xu} or polarized
optical spectroscopic techniques\cite{mci}, it in present becomes extremely
desirable to all-electrically detect the unique SML nature and resulting
spin texture from electron transports. Theoretically, two-terminal spin valve%
\cite{Yoko,Tagu} and three-terminal asymmetric structure\cite{Roy,Saha,Das}
have been suggested to extract the information of SML by probing the
spin-polarized currents along TI surface. In realistic experiments\cite%
{liu,Dank,tia,li}, however, it remains great challenging due to unavoidable
disturbance from bulk states\cite{zhou}.

Alternatively, the imaging of scanning tunneling microscopy (STM) serves as
a powerful tool to probe the nature of topological surface states by
analyzing the quasi-particle interference (QPI) in Fourier-transform
scanning-tunneling spectroscopy\cite{Rous,son,oka,alp,zhang2,zhou,yos,str},
caused by scattering off impurities on the TI surface. The SML nature is
manifested indirectly by the absence of backscattering between states of
opposite momentum and opposite spin. Nevertheless, these QPI patterns do not
show any signature of magnetic scattering even if the forbidden
backscattering is lifted since the QPI reveals only the spin-conserving
scattering. To extract the fingerprint of spin texture, the measurement of
magnetization patterns with spin-polarized STM was suggested\cite%
{zhou,str,kal,wang2,Liu2}. One, however, can note that most experiments only
focus on the probing of the out-of-plane spin texture\cite{yan,oka2} while
the in-plane spin texture, vital for understand the SML nature, receives no
attention due to complex physics in TIs. According to theory\cite%
{wor,ter,wie}, the spin-resolved STM conductance $dI/dV$ links to the
magnetization of tip and sample through
\begin{equation}
dI(\mathbf{r})/dV\propto \rho _{t}\rho (\mathbf{r},eV)+\left\vert \mathbf{m}%
_{t}\right\vert \left\vert \mathbf{M}(\mathbf{r},eV)\right\vert \cos \theta .
\end{equation}%
Here, $\rho _{t}[\rho (\mathbf{r},eV)]$ and $\mathbf{m}_{t}[\mathbf{M}(%
\mathbf{r},eV)]$ are, respectively, the charge and magnetization density of
tip (substrate), and $\theta $ is the angle between the tip and sample
magnetization. In Eq. (1), if polarizing the substrate $\mathbf{M}(\mathbf{r%
},eV)=M_{z}(\mathbf{r},eV)\hat{z}$ along z-direction, perpendicular to the
surface, the conductance is proportional only to the polar angle $\theta
_{t} $ of tip magnetization, but independent of its azimuthal angle $\varphi
_{t}$. The situation, however, is radically changed if the substrate is the
polarized TIs since the spin polarization $M_{z}(\mathbf{r},eV)$ can induce
the extra in-plane components $\mathbf{M}_{\parallel }(\mathbf{r},eV)=(M_{x}(%
\mathbf{r},eV), M_{y}(\mathbf{r},eV))$ due to strong spin-orbit interactions%
\cite{chi,bis}. As a consequence, the total magnetization distorts from the
primary z-axis and contributes an extra component of conductance. In this
paper, we get insight into this new physics and modify the formula (1) to be
suitable for the helical topological surface. It is found that the
conductance is dependent on the azimuthal angle of tip magnetization, which
has no analog in conventional magnetic metal, from which one can
electrically probe the in-plane spin texture in real space and further to
extract the SML angle of pristine topological states.

\begin{figure}[tbp]
\centering \includegraphics[width=0.35\textwidth]{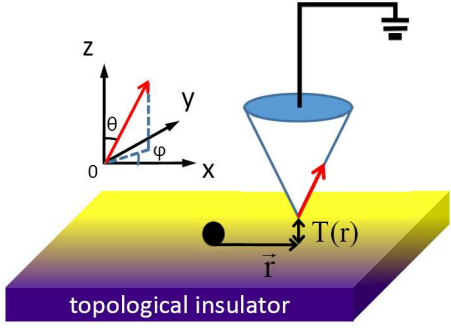}
\caption{(Color online) Schematic representation of the proposed
experimental setup. The polar ($\protect\theta_{t}$) and the azimuthal
angles($\protect\varphi _{t}$) of the tip magnetization $\mathbf{m_{t}}$ are
shown. All the in-plane angles are measured with respect to the direction of
x axis anticlockwise. $\mathbf{r}$ is in-plane vector of tip position
measured from impurity point and $T(\mathbf{r})$ denotes the tip-surface
coupling.}
\end{figure}
\emph{Formulas for spin-polarized transports--} To connect $\mathbf{M}%
_{\parallel}(\mathbf{r},eV)$ to conductance, we employ a typical
experimental setup as shown in Fig. 1, where a spin-polarized STM tip is
placed over a host surface of TIs, absorbed by a magnetic impurity whose
position is chosen as the original point $(\mathbf{r}=0)$. The introduction
of magnetic impurity has threefold meanings: (1) Polarizing the topological
surface. Although the topological surface states have a specific spin
orientation in momentum space, they have no net polarization in real space
due to the presence of time-reversal symmetry; (2) Inducing the planar
magnetism $\mathbf{M}_{\parallel}(\mathbf{r},eV)$ and (3) generating $%
\varphi _{t}$-dependent current. The points (2) and (3) are based on
breaking of the spatial rotate symmetry by the impurity. For undoped
surface, the in-plane magnetism and $\varphi _{t}$-dependent current must
vanish when integrating over the momentum.
We model the Hamiltonian of the spin-polarized STM tip as $H_{tip}=\sum_{%
\mathbf{k}}c_{t\mathbf{k}}^{\dagger }\left[ \epsilon _{\mathbf{k}}+\mathbf{m}%
_{t}\cdot \mathbf{\sigma }\right] c_{t\mathbf{k}}$, with $\mathbf{m}_{t}\ $%
the magnetization vector and $\mathbf{\sigma }$ the vector of spin Pauli
matrices, and the hybridized Hamiltonian between tip and topological surface
as $H_{hyb}=\int \int d\mathbf{r}_{1}d\mathbf{r}_{2}\psi _{t}^{\dagger }(%
\mathbf{r}_{1},t)T(\mathbf{r}_{1},\mathbf{r}_{2})\psi _{s}\left( \mathbf{r}%
_{2},t\right) +h.c.,$ where the quantum field operators $\psi _{\eta
}^{\dagger }\left( \mathbf{r,t}\right) =\frac{1}{\sqrt{N}}\sum_{\mathbf{k}}$
$c_{\eta \mathbf{k}}^{\dagger }(t)e^{-i\mathbf{k}\cdot \mathbf{r}}$ with $%
c_{\eta \mathbf{k}}^{\dagger }(t)=(c_{\eta \mathbf{k\uparrow }}^{\dagger
}(t),c_{\eta \mathbf{k\downarrow }}^{\dagger }(t))$ is the creation operator
of elections for surface ($\eta =s$) and tip ($\eta =t)$. We choose the
spin-quantization axis of the surface electrons as the global reference
axis. The tip-surface coupling is assumed to be spin independent $T(\mathbf{r%
}_{1},\mathbf{r}_{2})=T_{0}(\mathbf{r}_{1})\delta (\mathbf{r}_{2}-\mathbf{r}%
) $ with $\mathbf{r}$ being the in-plane spatial vector of tip measured from
the impurity point. Here, the coupling between the tip and impurity is
neglected due to weak interaction since we focus on large $\mathbf{r}$. By
introduction of unitary matrix
\begin{equation}
U=\left[
\begin{array}{cc}
\cos \left( \theta _{t}/2\right) e^{-i\varphi _{t}} & -\sin \left( \theta
_{t}/2\right) e^{-i\varphi _{t}}\\
\sin \left( \theta _{t}/2\right) & \cos \left( \theta _{t}/2\right)%
\end{array}
\right],
\end{equation}

where $\theta _{t}$($\varphi _{t}$) is the polar (azimuthal) angle of the %
tip magnetization $\mathbf{m}_{t}$, one can diagonalize $H_{tip}=\sum_{%
\mathbf{k}\alpha }(\epsilon _{\mathbf{k}}+\alpha \left\vert \mathbf{m}%
_{t}\right\vert )\gamma _{t,\mathbf{k}\alpha }^{\dagger }\gamma _{t,\mathbf{k%
}\alpha }$. Here, the quasi-particle operator $(\gamma _{\mathbf{k+}},
\gamma_{\mathbf{k-}})=U^{\dagger }(c_{\mathbf{k\uparrow }},c_{t,\mathbf{%
k\downarrow }})$ with $\alpha =\pm $ parallel (antiparallel) to tip%
magnetization. In this basis, the spin flipping due to noncollinear%
arrangement between the magnetic moments of the substrate and tip enters the%
tip-surface tunneling, which is rewritten as
\begin{equation}
H_{hyb}=\int \int d\mathbf{r}_{1}d\mathbf{r}_{2}\psi _{t}^{\dagger }(\mathbf{%
r}_{1},t)\tilde{T}(\mathbf{r}_{1},\mathbf{r}_{2})\gamma _{t}\left( \mathbf{r}%
_{2},t\right) +h.c.,
\end{equation}

where the renormalized coupling matrix $\tilde{T}(\mathbf{r}_{1},\mathbf{r}%
_{2})=T(\mathbf{r}_{1},\mathbf{r}_{2})U^{-1}$ has nondiagonal term in spin space.
The current through the tip is calculated with $I=-e\frac{\partial }%
{\partial t}\sum_{\alpha }\int d\mathbf{r}_{1}\times\langle \Psi _{t}^{\alpha
\dagger }\left( \mathbf{r}_{1},t\right) \Psi _{t}^{\alpha }\left( \mathbf{r}%
_{1},t\right) \rangle $. Carrying out the equation of motion for
non-equilibrium Green's function on the Keldysh technique, we obtain the
conductance as (see the Supplementary Methods)

\begin{eqnarray}
I&=&-2e\left\vert T_{0}\right\vert ^{2}\sum_{\mathbf{p}}\int \frac{d\omega }{%
2\pi }\notag \\
&&\times Tr\{\text{Re}[g^{r}(\mathbf{r},\mathbf{r},\omega )Ug_{\mathbf{p}%
}^{<}(\omega )U^{-1}\notag  \\
&&+g^{<}(\mathbf{r},\mathbf{r,}\omega )Ug_{t\mathbf{p}}^{a}(\omega )U^{-1}]\},
\end{eqnarray}

where $g^{r(<)}(\mathbf{r},\mathbf{r};\omega )$ is the retarded (lesser)
Green's function of topological surface states in real-frequency space and $%
g_{\mathbf{p}}^{a(<)}(\omega )$ is the advanced (lesser) Green's function of
tip in momentum-frequency space. We denote $T_{0}=\int d\mathbf{r}^{\prime
}e^{-i\mathbf{p}\cdot \mathbf{r}^{\prime }}T_{0}(\mathbf{r}^{\prime })$
assumed independence of momentum. Compared to previous derivation\cite%
{pent,fra,rui2}, an important difference is the matrix $g^{r}(\mathbf{r},%
\mathbf{r};\omega )$ including the spin flipping when Dirac electrons travel
on the topological surface.

As usual, we define the charge density of TIs as $\rho (\mathbf{r},\omega )=-%
\frac{1}{2\pi }Im$Tr$[g(\mathbf{r},\mathbf{r},\omega +i0^{+})]$ and its spin
texture as $\mathbf{M}(\mathbf{r},\omega )=-\frac{1}{2\pi }ImTr[\frac{%
\mathbf{\sigma }}{2}g(\mathbf{r},\mathbf{r},\omega +i0^{+})]$. Finally, we
in the zero-temperature limit obtain the expression for conductance at bias $%
eV$, which can be divided to two parties $G(\mathbf{r})=G_{0}(\mathbf{r}%
)+G_{flip}(\mathbf{r})$, with%

\begin{eqnarray}
G_{0}(\mathbf{r}) &=&\pi e\left\vert T_{0}\right\vert ^{2}[\rho (\mathbf{r}%
,eV)\rho _{t} \notag\\
&&+M_{z}(\mathbf{r},eV)\left\vert \mathbf{m}_{t}\right\vert \cos
\theta _{t}], \\
G_{flip}(\mathbf{r}) &=&\pi e\left\vert T_{0}\right\vert ^{2}\left\vert
\mathbf{m}_{t}\right\vert \left\vert \mathbf{M}_{\parallel }(\mathbf{r}%
,eV)\right\vert \notag\\
&&\times \sin \theta _{t}\cos (\varphi _{t}-\varphi _{\mathbf{M}}),
\end{eqnarray}%
where $\varphi _{\mathbf{M}}$ is the azimuthal angle of $\mathbf{M}(\mathbf{r%
},eV)$. The conductance $G_{0}(\mathbf{r})$ recovers the usual formula in
Eq. (1), which is azimuthal independent. The most interest is $G_{flip}(%
\mathbf{r})$ in Eq. (6), which is contributed by spin-flipping process
contained in Green's function $g_{\mathbf{\sigma \bar{\sigma}}}^{r}(\mathbf{r%
},\mathbf{r},\omega )$ when an electron is scattered off the magnetic
impurity. Importantly, such dependence of the tunneling conductance on the
azimuthal angle of the tip magnetization has no analog in conventional
magnetic metals.

\emph{Probing of spin texture in linear dispersion--} In this section, we
will demonstrate that how the measurement of $G_{flip}(\mathbf{r})$ with a
spin-polarized STM reconstructs the spin texture of the TI surface states
and then further determines its SML angle in the real space. It is easy to
verify that the presence of magnetic impurities is strictly necessary since $%
M_{\parallel }(\mathbf{r},eV)$ vanishes without the magnetic impurity.

To calculate $\mathbf{M}_{\parallel }(\mathbf{r},eV)$ in Eq. (6), we must
first obtain the dirtied Green's function of Dirac electrons $g(\mathbf{r},%
\mathbf{r},\omega )$, which can be calculated with T-matrix approach\cite%
{bal,mah,rui}.
\begin{equation} \label{Grr}
g\left( \mathbf{r},\mathbf{r},\omega \right) =g_{0}\left( 0,\omega \right)
+g_{0}\left( \mathbf{r},\omega \right) T\left( \omega \right) g_{0}\left( -%
\mathbf{r},\omega \right) ,
\end{equation}%
which takes into account the multiple scattering events of electrons by the
impurity. Here, the impurity-free Green's function $g_{0}\left( \mathbf{r}%
,\omega \right) $ is the Fourier transform of $g_{0}\left( \mathbf{k},\omega
\right) =\left[ \omega +i0^{+}-H_{TI}^{0}\right] ^{-1}$ with respect to the
bare TI Hamiltonian $H_{TI}^{0}$, and the T-matrix is given by the
Bethe-Scalpeter equation $T\left( \omega \right) =V\left[ 1-Vg_{0}(0,\omega )%
\right] ^{-1}$. The impurity potential is assumed in the form of $V=\left(
U_{0}-U_{z}\sigma _{z}\right) $, consisting of a scalar potential $U_{0}$
and a magnetic potential $U_{z}$ polarized perpendicular to the surface. The
Hamiltonian of surface of TIs is described by
\begin{equation}
H_{TI}^{0}(\lambda )=\sum_{\mathbf{k}}[\hbar v_{f}(\mathbf{\sigma \times k}%
)_{z}+\frac{\lambda }{2}(k_{+}^{3}+k_{-}^{3})\sigma _{z}],
\end{equation}%
which includes the warping term with strength $\lambda $ and $k_{\pm
}=k_{x}\pm ik_{y}$.

\begin{figure}[tbp]
\centering \includegraphics[width=0.48\textwidth]{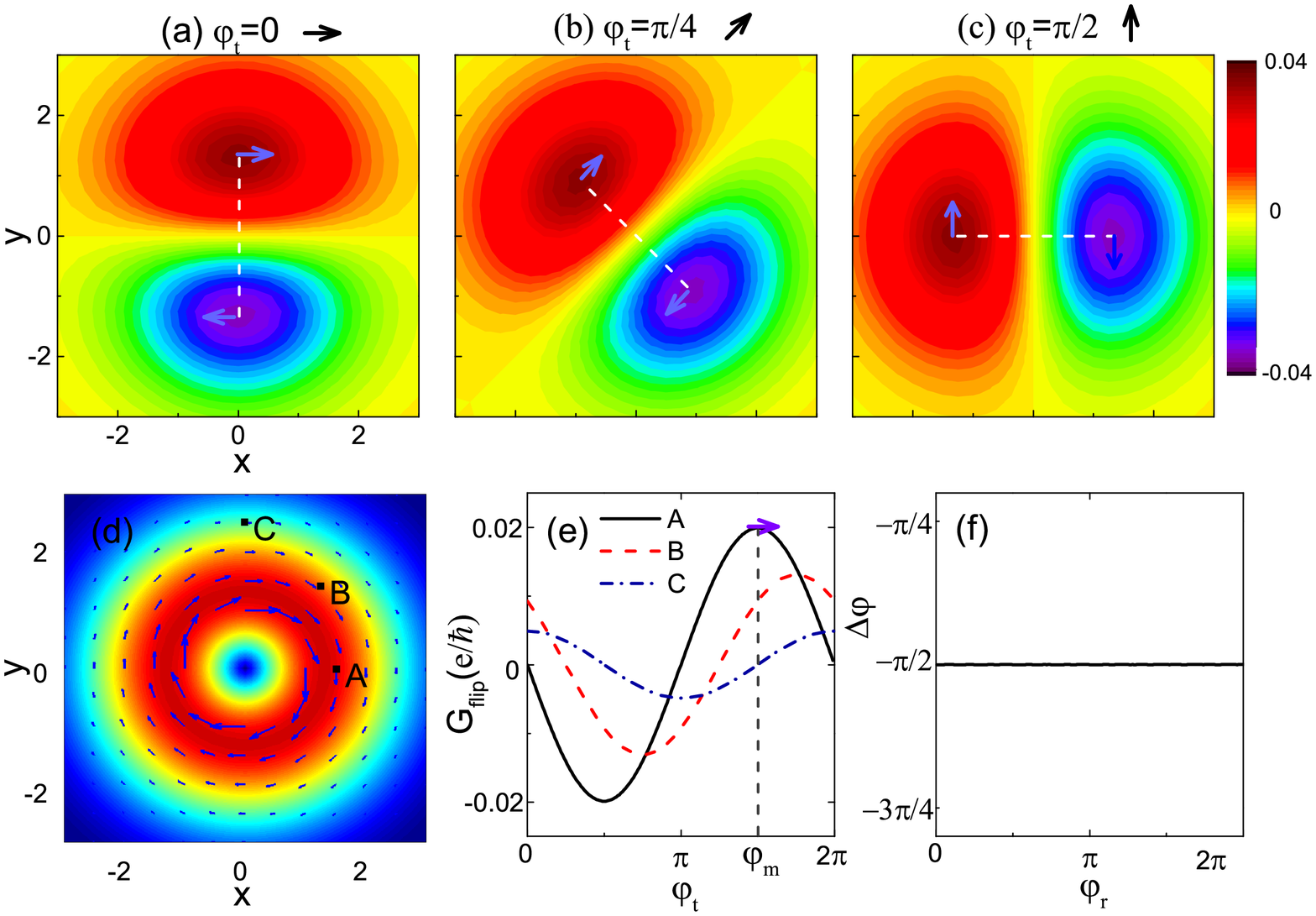}
\caption{(Color online) (a)-(c) Spatially resolved conductance maps for
different directions of tip magnetization $\protect\varphi_{t}=0,\protect\pi %
/4,\protect\pi/2 $ with $\theta_{t}=\pi/2$ and $eV=0.2$. (d) The distribution of in-plane spin texture $\mathbf{M}%
_{\parallel }(\mathbf{r},eV)$ with arrows showing the directions and the
color code representing the degree of the spin polarization. (e) Dependence
of $G_{flip}(\mathbf{r})$ on $\protect\varphi_{t}$ for three points A($r=1.5$ and $\varphi=0$),%
 B($r=2$ and $\varphi=\pi/4$), and C($r=2.5$ and $\varphi=\pi/2$) indicated in (d).%
(f) Direction difference $\Delta\protect\varphi%
=\left\vert\protect\varphi _{\mathbf{r}}-\protect\varphi _{\mathbf{M}%
}\right\vert$ between $\mathbf{r}$ and $\mathbf{M}_{\parallel }(\mathbf{r}%
,eV)$ as a function of spatial direction $\protect\varphi _{\mathbf{r}}$ for $r=2$.
The chosen other parameters are $D_{c}=1.5,$ $\hbar=1,$ $v_{f}=1,$ %
$U_{0}=0,$ $U_{z}=100,$ $|\mathbf{m}_{t}|=0.5,$ $\rho_{t}=1,$ and $T_{0}=1$ for all diagrams.}
\end{figure}
We first consider the case of $\lambda =0$ and plot the real-space
distribution of the in-plane spin texture $\mathbf{M}_{\parallel }(\mathbf{r}%
,eV)$ in Fig. 2(d) and of the spin-flipping conductance $G_{flip}(\mathbf{r}%
) $ in Figs. 2(a)-(c) for different azimuthal angles $\varphi_{t}$ of the
polarized tip. For a fixed tip direction $\varphi_{t}$, $G_{flip}(\mathbf{r}%
) $ is spatial asymmetry, with the extremum in a circle around the original
point appearing along certain radius (dashed white line). The positive and
negative maxima, respectively, correspond to the in-plane magnetization $%
\mathbf{M}_{\parallel }(\mathbf{r},eV)$ parallel and antiparallel to the
polarized direction of the STM tip, as indicated by arrows, due to spin
selection of the tip. As one rotates the tip direction $\varphi_{t}=0,\pi
/4,\pi/2 $, the extremum position also rotates anticlockwise with the
unchanged magnitude, indicating the in-plane magnetization $\mathbf{M}%
_{\parallel }(\mathbf{r},eV)$ tangential to the concentric circle with
clockwise helicity. To accurately determine the orientation of $\mathbf{M}%
_{\parallel }(\mathbf{r},eV)$, we depict the dependence of $G_{flip}(\mathbf{%
r})$ on $\varphi _{t}$ in Fig. 2(e), where the position of peak just
corresponds to $\varphi _{\mathbf{M}}$. By choosing any different positions,
e.g., points A, B, and C indicated in Fig. 2(d), it is found that the
direction between $\mathbf{r}$ and $\mathbf{M}_{\parallel }(\mathbf{r},eV)$
always satisfies a relation $\Delta\varphi=\left\vert\varphi _{\mathbf{r}%
}-\varphi _{\mathbf{M}}\right\vert=\pi /2$, where $\varphi _{\mathbf{r}}$ is
the azimuthal angle of $\mathbf{r}$, as shown in Fig. 2(f). Therefore, the
measurement of $G_{flip}(\mathbf{r})$ can ascertain the in-plane spin
texture in Fig. 2(d).
To understand the origin of relation $\left\vert\varphi _{\mathbf{r}%
}-\varphi _{\mathbf{M}}\right\vert=\pi /2$, we further derive the analytical
formula with $H_{TI}^{0}(\lambda =0)$, from which $g_{0}(0,\omega )=\frac{1}{%
4(\hbar v_{F})^{2}}[\frac{\omega }{\pi }\ln (\frac{\omega ^{2}}{%
D_{c}^{2}-\omega ^{2}})-i\left\vert \omega \right\vert \Theta \left(
D_{c}-\left\vert \omega \right\vert \right) ]$ and
\begin{equation}
g_{0}(\mathbf{r},\omega )=-\frac{\omega }{2\pi v_{F}^{2}}\left(
\begin{array}{cc}
K_{0}(\xi ) & e^{-i\varphi _{\mathbf{r}}}K_{1}(\xi ) \\
e^{i\varphi _{\mathbf{r}}}K_{1}(\xi ) & K_{0}(\xi )%
\end{array}%
\right).
\end{equation}%
Here, $D_{c}$ is the cutoff energy for the band width of surface states, $%
\xi =-i\left\vert\mathbf{r}\right\vert\omega /\hbar v_{F}$, and $K_{n}(x)$
is the Bessel functions of the $n$th-order. $\varphi _{\mathbf{r}}$ in Eq.
(9) arises from the Fourier transform of momentum direction $\varphi _{%
\mathbf{k}}$. Finally, we obtain azimuthal angle of the in-plane
magnetization
\begin{equation}
\cos \varphi _{\mathbf{M}}=M_{x}/\sqrt{M_{x}^{2}+M_{y}^{2}}=\sin \varphi _{%
\mathbf{r}},
\end{equation}%
and its magnitude $\left\vert \mathbf{M}_{\parallel }(\mathbf{r}%
,eV)\right\vert =\frac{\omega ^{2}U_{z}}{A\pi ^{3}v_{F}^{4}}K_{0}(\xi
)K_{1}(\xi )$ with $A=1-2g_{0}(0,\omega )U_{0}-g_{0}(0,\omega
)^{2}(U_{z}^{2}-U_{0}^{2})$. Thus, we obtain $\left\vert \varphi _{\mathbf{M}%
}-\varphi _{\mathbf{r}}\right\vert =\pi /2$, which is corresponding to SML
angle $\left\vert \varphi _{\mathbf{M}}-\varphi _{\mathbf{k}}\right\vert
=\pi /2$ of the pristine TI in momentum space. Notice that though the
in-plane magnetization in real space is induced by the impurity magnetism,
it also stems purely from the spin-orbit effect. Consequently, the planar $%
\mathbf{M}_{\parallel }(\mathbf{r},eV)$ still contains the information of
SML, reflected by the spin-position locking.

With Eq. (10), we can further rewrite Eq. (6) as
\begin{equation}
\mathbf{M}_{\parallel }(\mathbf{r},eV)=\left\vert \mathbf{M}_{\parallel }(%
\mathbf{r},eV)\right\vert \mathbf{\hat{r}}\times \hat{U}_{z},
\end{equation}%
where $\hat{U}_{z}$ is the unit vector along z direction. Obviously, $%
\mathbf{M}_{\parallel }(\mathbf{r},eV)$ is perpendicular to $\mathbf{r}$,
which is a consequence of the Dzyaloshinskii-Moriya interaction caused by
the helical spin structure of Dirac surface states on TIs. Therefore, the
measurement of $\mathbf{M}_{\parallel }(\mathbf{r},eV)$ not only provides a
way to extract the pristine SML angle, but also the strength of the DM
interaction. In addition, here due to $G_{flip}(\mathbf{r})\propto \cos
(\varphi _{t}-\varphi _{\mathbf{r}}+\pi /2)$, rotating the tip has equal
role with the rotation of spatial position around the impurity, which is
helpful in realistic measurement.

As for the magnitude of $\mathbf{M}_{\parallel }(\mathbf{r},eV)$ at certain
point $\mathbf{r}$, we can determine it by rotation of tip direction,
namely, $\left\vert \mathbf{M}_{\parallel }(\mathbf{r},eV)\right\vert ={%
Max[G_{flip}(\mathbf{r,}\varphi _{t})]-Min[G_{flip}(\mathbf{r,}\varphi
_{t}^{\prime })]}/B$ with $B=2\pi e\left\vert T_{0}\right\vert
^{2}\left\vert \mathbf{m}_{t}\right\vert $.

\begin{figure}[tbp]
\centering \includegraphics[width=0.48\textwidth]{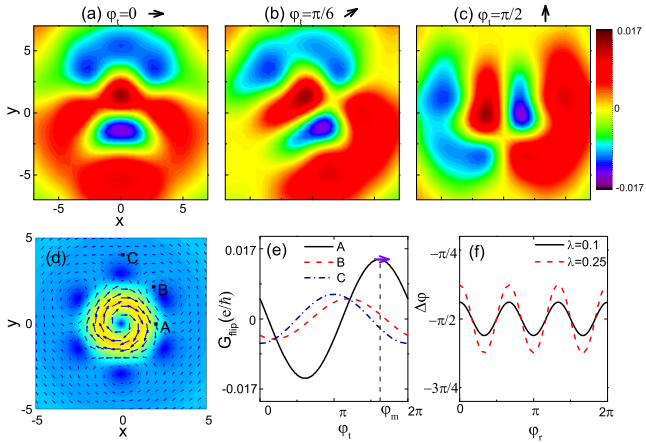}
\caption{(Color online) (a)-(c) Spatially resolved conductance in warping dispersion $\lambda=0.25$ maps for
different directions of tip magnetization $\protect\varphi_{t}=0,\protect\pi %
/6,\protect\pi/2 $ with $eV=0.6$. (d) The distribution of in-plane spin texture $\mathbf{M}%
_{\parallel }(\mathbf{r},eV)$ with arrows showing the directions and the
color code representing the degree of the spin polarization. (e) Dependence
of $G_{flip}(\mathbf{r})$ on $\protect\varphi_{t}$ for three points A ($r=2$ and $\varphi=0$), B($r=3$ and $\varphi=\pi/4$),
and C($r=4$ and $\varphi=\pi/2$) indicated in (d). (f) Direction difference $\Delta\protect\varphi$ as
a function of spatial direction $\protect\varphi _{\mathbf{r}}$ for $r=2$. The rest of parameters refer to data used above.}
\end{figure}
\emph{Probing of spin texture in warping dispersion--} For $\lambda \neq 0$,
the magnetization density $\mathbf{M}_{\parallel }(\mathbf{r},eV)$ is
demonstrated in Fig. 3(d). Compared to Fig. 2(d), introduction of the
warping term greatly modifies the surface magnetism, i.e., not only
modifying $M_{z}(\mathbf{r},eV)$ but also making $\mathbf{M}_{\parallel }(%
\mathbf{r},eV)$ deviate from the circular structure or $\left\vert \varphi _{%
\mathbf{M}}-\varphi _{\mathbf{r}}\right\vert \neq \pi /2$. Especially, for
large distance $\left\vert \mathbf{r}\right\vert$, there appear three new
circular centers with anticlockwise helical spin structure, exhibiting $C_3$
symmetry of lattice structure describing by the TI Hamiltonian in Eq. (8). A
main reason for large change of the spin texture is that the warping term
generates additional in-plane magnetization $M_{r}(\mathbf{r},eV)$ along
radial direction(or $\mathbf{r}$). The component $M_{r}(\mathbf{r},eV)$ is
attributed to the antiferromagnetic Ruderman-Kittel-Kasuya-Yosida
interaction along the line joining the impurities\cite{bis}. In this
situation, $\mathbf{M}_{\parallel }(\mathbf{r},eV)$ has a complex dependence
on the spatial direction. When we scan the tip over the whole surface with
fixed tip azimuthal angle, e.g., $\varphi _{t}=0$ in Fig. 3(a), the
alternating positive and negative maxima of conductance along radial
direction reflect the alternating change of spin structure in Fig. 3(e).
With the tip rotating from Figs. 3(a) to (c), the extremum also rotates
anticlockwise. Unlike the case without warping term in Figs. 2(a)-(c), the
structure of extremum regime from Figs. 3(a) to (c) is changed, indicating
the spin texture deviating from the concentric circle. Even so, we still can
exactly determine the direction of magnetization at arbitrary point only by
rotating the tip direction around the z axis. For example, to determine the
direction $\varphi _{\mathbf{M}}$ of magnetization in points A, B, and C
labeled in Fig. 3(d), one can plot $G_{flip}(\mathbf{r})$ vs $\varphi _{t}$
as shown in Fig. 3(e). $\varphi _{\mathbf{M}}$ is equal to the size of $%
\varphi _{t}$ at the conductance peak.

\begin{figure}[tbp]
\centering \includegraphics[width=0.45\textwidth]{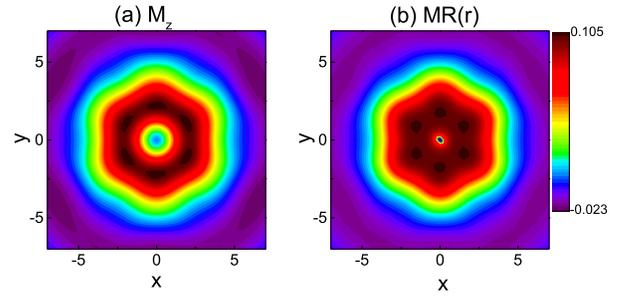}
\caption{(Color online) Spatially resolved out-of-plane spin texture $%
\mathbf{M}_{z}(\mathbf{r},eV)$ (a) and magnetoresistance $MR$ map (b) with
the color code representing the magnitude. The directions
of tip magnetization is $\theta_{t}=0$ with $eV=0.6$. And strength of warp term is $\lambda=1$. The rest of parameters refer to data used above.}
\end{figure}
From discussions in Fig. 2, we are known that the direction difference $%
\Delta\varphi=\left\vert\varphi _{\mathbf{r}}-\varphi _{\mathbf{M}%
}\right\vert$ between $\mathbf{r}$ and $\mathbf{M}_{\parallel }(\mathbf{r}%
,eV)$ can characterize the SML angle well. In Fig. 3(f), we depict $%
\Delta\varphi$ as a function of $\varphi _{\mathbf{r}}$ in order to clarify
the effect of warping term. As the warping term is added, $\Delta\varphi$
deviates from the perfect SML ($\Delta\varphi=\pi /2$) and exhibits an
oscillating behavior. With the increase of $\lambda $, the deviation
amplitude becomes larger and larger but at the same time the oscillating
period of $2\pi /3$ remains unchanged, reappearing the $C_{3}$ symmetry of
hexagonal lattice in Fig. 3(d). The change of $\varphi _{\mathbf{M}}$ is
remarkable along directions of $\varphi _{\mathbf{r}}=n\pi/3$ with $%
n=0,1,2,3,4,5$ due to the strong out-of-plane magnetization, which
corresponds to the center of the sides of the hexagon of the Fermi surface.
By contrast, for $\varphi _{\mathbf{r}}=n\pi /6$ with $n=1,3,5,7,9,11$
corresponding to the corners of the hexagonal pattern, the pristine perfect
SML angle $\Delta\varphi=\pi/2$ (or $\mathbf{M}_{\parallel }(\mathbf{r}%
,eV)\bot \mathbf{r}$) is still abided by due to the vanishing out-of-plane
magnetization.

An important feature for the TI with warping term is the staggered structure
of the out-of-plane magnetization $M_{z}(\mathbf{r},eV)$ as shown in Fig.
4(a), where alternating positive and negative values have six symmetric
regions. To probe its complex spin texture, we can set $\theta _{t}=0$ and
so the total conductance is $G_{0}(\mathbf{r})$, whose spin dependence of $%
G_{0}(\mathbf{r})$ stems completely from the $M_{z}(\mathbf{r},eV)$. In this
case, we keep the spin-polarized STM magnetization either parallel or
antiparallel to the z axis to define the magnetoresistance (MR) effect,
given by
\begin{equation}
MR(\mathbf{r})=\frac{G_{0}(\mathbf{r},\varphi _{t}=0)-G_{0}(\mathbf{r}%
,\varphi _{t}=\pi )}{G_{0}(\mathbf{r},\varphi_{t}=0)+G_{0}(\mathbf{r}%
,\varphi _{t}=\pi )}.
\end{equation}
We depict $MR(\mathbf{r})$ in Fig. 4(b), which exhibits six regions with
alternating high and low conductance density, completely reconstituting the
spatial pattern of spin texture $M_{z}(\mathbf{r},eV)$ in Figs. 4(a).

\emph{Conclusions--} In conclusion, employing the nonequilibrium Green's
function, we present a general theory for spin polarized transports of Dirac
electrons through a spin-polarized STM. In order to extract quantitative
information about spin texture, we need to break the symmetry of rotation
and time reversal with a typical impurity model. It is found that the
conductance is modified by an extra component exclusively associated with in
xy-plane spin texture, which exhibits an unconventional dependence on the
azimuthal angle of the tip magnetization. The analysis of azimuthal angle
dependent conductance provides a direct method of measurement of the local
in-plane spin texture of the Dirac electrons on the surface. By measurement
of the spatial resolved conductance $G_{flip}(\mathbf{r})$, we can not only
reconstruct the helical structure of spin texture but also can extract the
SML angle if the in-plane magnetization is induced purely by the spin-orbit
coupling of the surface Dirac election. Experimentally, the magnetic tip can
be prepared by coating antiferromagnetic Cr on tungsten tips where the
magnetic direction of the tip can be controlled by the Cr thickness, either
in plane ($>30$nm) or out of plane ($\sim5$ nm)\cite{wac,wie}. Therefore,
the measurement of the in-plane magnetization offers an alternative way to
identify the topology nature on TI surfaces.

\section{ACKOWLEDGMENTS}

This work was supported by National Natural Science Foundation of China
(Grant Nos. 11474106 and 11274124), and by the Innovation Project of
Graduate School of South China Normal University.

\section{Authors contributions}

R.Q.W. conceived the idea. S.H.Z. performed the calculation and provided all
of the figures. R.Q.W. and S.H.Z. contributed to the interpretation of the
results and wrote the manuscript. H.J.D. and M.Y. joined in the data
analysis and contributed in the discussion. All authors reviewed the
manuscript.

\section{Additional Information}

\textbf{Competing financial interests:} The authors declare no competing
financial interests.

\textbf{Figure legends} Fig.1: Schematic representation of the proposed
experimental setup. The polar ($\protect\theta_{t}$) and the azimuthal
angles($\protect\varphi _{t}$) of the tip magnetization $\mathbf{m_{t}}$ are
shown. All the in-plane angles are measured with respect to the direction of
x axis anticlockwise. $\mathbf{r}$ is in-plane vector of tip position
measured from impurity point and $T(\mathbf{r})$ denotes the tip-surface
coupling.

Fig.2: (a)-(c) Spatially resolved conductance maps for
different directions of tip magnetization $\protect\varphi_{t}=0,\protect\pi %
/4,\protect\pi/2 $ with $\theta_{t}=\pi/2$ and $eV=0.2$. (d) The distribution of in-plane spin texture $\mathbf{M}%
_{\parallel }(\mathbf{r},eV)$ with arrows showing the directions and the
color code representing the degree of the spin polarization. (e) Dependence
of $G_{flip}(\mathbf{r})$ on $\protect\varphi_{t}$ for three points A($r=1.5$ and $\varphi=0$),%
 B($r=2$ and $\varphi=\pi/4$), and C($r=2.5$ and $\varphi=\pi/2$) indicated in (d).%
(f) Direction difference $\Delta\protect\varphi%
=\left\vert\protect\varphi _{\mathbf{r}}-\protect\varphi _{\mathbf{M}%
}\right\vert$ between $\mathbf{r}$ and $\mathbf{M}_{\parallel }(\mathbf{r}%
,eV)$ as a function of spatial direction $\protect\varphi _{\mathbf{r}}$ for $r=2$.
The chosen other parameters are $D_{c}=1.5,$ $\hbar=1,$ $v_{f}=1,$ %
$U_{0}=0,$ $U_{z}=100,$ $|\mathbf{m}_{t}|=0.5,$ $\rho_{t}=1,$ and $T_{0}=1$ for all diagrams.

Fig.3: (a)-(c) Spatially resolved conductance in warping dispersion $\lambda=0.25$ maps for
different directions of tip magnetization $\protect\varphi_{t}=0,\protect\pi %
/6,\protect\pi/2 $ with $eV=0.6$. (d) The distribution of in-plane spin texture $\mathbf{M}%
_{\parallel }(\mathbf{r},eV)$ with arrows showing the directions and the
color code representing the degree of the spin polarization. (e) Dependence
of $G_{flip}(\mathbf{r})$ on $\protect\varphi_{t}$ for three points A ($r=2$ and $\varphi=0$), B($r=3$ and $\varphi=\pi/4$),
and C($r=4$ and $\varphi=\pi/2$) indicated in (d). (f) Direction difference $\Delta\protect\varphi$ as
a function of spatial direction $\protect\varphi _{\mathbf{r}}$ for $r=2$. The rest of parameters refer to data used above.

Fig.4: Spatially resolved out-of-plane spin texture $%
\mathbf{M}_{z}(\mathbf{r},eV)$ (a) and magnetoresistance $MR$ map (b) with
the color code representing the magnitude. The directions
of tip magnetization is $\theta_{t}=0$ with $eV=0.6$. And strength of warp term is $\lambda=1$. The rest of parameters refer to data used above.

\end{document}